\documentclass[10pt,a4paper,notitlepage,prd,showpacs,preprintnumbers,superscriptaddress,nofootinbib,twocolumn]{revtex4-1}
\usepackage{float}
\usepackage{amsmath,amssymb}
\usepackage{cancel}
\usepackage{graphicx}

\makeatletter

\usepackage{bm}
\usepackage{tensor}
\usepackage{comment}
\usepackage{ifpdf}
\usepackage{slashed}
\usepackage{color}
\usepackage[mathscr]{eucal}

\ifpdf
  \usepackage{graphicx}     
  \usepackage[bookmarksopen,colorlinks=true,linkcolor=bblue,citecolor=bblue,urlcolor=ppink]{hyperref}
\else     
  \usepackage[dvipdfmx]{}     
  \usepackage[dvipdfmx,bookmarksopen,colorlinks=true,linkcolor=bblue,citecolor=ppink,urlcolor=darkred]{hyperref}
\fi

\definecolor{red}{rgb}{1,0,0}
\definecolor{darkred}{rgb}{0.6,0,0}
\definecolor{darkgreen}{rgb}{0.992447,0.623778,0.034597}
\definecolor{ppink}{rgb}{1,0.4,0.4}
\definecolor{bblue}{rgb}{0.284602,0.317763,0.963947}

\newcommand{\Mpl}{M_{\rm Pl}}

\def\Mpl{M_{\rm Pl}}
\def \hc{\rm{h.c.}}

\newcommand{\ltsim}{\protect\raisebox{-0.5ex}{$\:\stackrel{\textstyle <}{\sim}\:$}}
\newcommand{\gtsim}{\protect\raisebox{-0.5ex}{$\:\stackrel{\textstyle >}{\sim}\:$}}

\newcommand{\footnoteref}[1]{\protected@xdef\@thefnmark{\ref{#1}}\@footnotemark}

\makeatother

\begin{document}

\preprint{IPMU17-0151}

\title{Cogenesis of LIGO Primordial Black Holes and Dark Matter}

\author{Fuminori Hasegawa}
\email[fuminori@icrr.u-tokyo.ac.jp]{}
\affiliation{ICRR, University of Tokyo, Kashiwa, 277-8582, Japan}

\affiliation{Kavli IPMU (WPI), UTIAS, University of Tokyo, Kashiwa, 277-8583,
Japan}

\author{Masahiro Kawasaki}
\email[kawasaki@icrr.u-tokyo.ac.jp]{}
\affiliation{ICRR, University of Tokyo, Kashiwa, 277-8582, Japan}

\affiliation{Kavli IPMU (WPI), UTIAS, University of Tokyo, Kashiwa, 277-8583,
Japan}
\begin{abstract}
\noindent In this letter, we propose a novel scenario which simultaneously explains $\mathcal{O}(10)M_\odot$ primordial black holes~(PBHs) and dark matter in the minimally supersymmetric standard model. 
Gravitational waves (GWs) events detected by LIGO-Virgo collaboration suggest an existence of black holes as heavy as $\sim 30M_\odot$. 
In our scenario, as seeds of the PBHs, we make use of the baryon number perturbations which are induced by the special type of Affleck-Dine mechanism.
Furthermore, the scenario does not suffer from the stringent constraints from CMB $\mu$-distortion due to the Silk damping and pulsar timing. 
We find the scenario can explain not only the current GWs events consistently, but also dark matter abundance by the non-topological solitons formed after Affleck-Dine mechanism, called Q-balls. 
\end{abstract}
\maketitle

\section{Introduction}
The LIGO-Virgo collaboration has announced the detection of the four GW events, GW150914~\cite{TheLIGOScientificCollaboration2016b}, GW151226~\cite{TheLIGOScientificCollaboration2016a}, GW170104~\cite{TheLIGOScientificCollaboration2017a}, GW170814~\cite{Abbott2017}. 
These events come from mergers of binary black holes~(BHs). However, among the observed eight BHs, four BHs have a mass $\sim 30M_\odot$.
There are disputes about formation of such heavy BH binaries by stellar evolution and many researchers are exploring the origin of those BHs.

Primordial black holes (PBHs) are one of the candidates which account for these GW events~\cite{Bird2016,Clesse2016,Kashlinsky:2016sdv,Carr2016,Eroshenko2016,Sasaki2016}.
PBHs are formed by the gravitational collapse of the overdense Hubble patches in the early Universe. 
Therefore, on the contrary to the stellar ones, PBHs can have a very wide range of masses including $\sim 30M_\odot$. 
As an origin of large density perturbations required for PBH formation, inflation in the early universe is well-motivated and studied extensively~\cite{Yokoyama:1995ex,GarciaBellido:1996qt,Kawasaki:1997ju,Kawasaki2016,Inomata2016,Inomata2017aa,Inomata2017}. 
Since the density contrast generated by conventional inflation is predominantly scale invariant and too small to form the PBHs, much effort has been made to amplify the curvature perturbations only at the small scales. 

However, such amplified small scale perturbations are severely constrained by cosmological observations. 
First, they cause a distortion of the Cosmic Microwave Background~(CMB) due to the Silk damping. 
In fact, the observation of the $\mu$-distortion excludes the inflationary PBHs with mass $4\times10^2M_\odot\lesssim M_{\rm PBH}\lesssim4\times10^{13}M_\odot$ including supermassive BHs (SMBHs)~\cite{Kohri2014}.
Furthermore, large (scalar)~curvature perturbations source tensor perturbations by the second-order effect~\cite{Saito2008,Saito2009,Bugaev:2010bb}. 
The secondary GWs can be significantly larger than those of the first-order and constrained by observations of pulsar timing. 
The latest results of pulsar timing array~(PTA) experiments~\cite{Arzoumanian2015,Lentati2015,Shannon2015} exclude the inflationary PBHs with mass $0.1M_\odot\ltsim M_{\rm PBH}\ltsim10M_\odot$. 
Consequently, inflationary PBHs can explain the massive BHs only in limited mass range. 
Fortunately, there still exist some successful models of inflationary PBHs which can explain the LIGO events evading those constraints~\cite{Inomata2016,Inomata2017}.    

In this letter, we propose a novel scenario which explains the LIGO events evading all difficulties mentioned above in the minimal supersymmetric standard model~(MSSM). 
In the scenario, the Affleck-Dine~(AD) mechanism~\cite{Affleck1985,Dine1996} plays a crucial roles. 
We find that the more general choice of the coefficients of the Hubble induced mass for the AD-field realizes the generation of the spatially-inhomogeneous baryon asymmetry. 
As a result, some high-baryon regions produced by this mechanism become over-dense in the cosmological evolution and gravitationally collapse into PBHs.
Although the idea of the inhomogeneous baryogenesis itself was proposed by Dolgov $et~al.$\cite{Dolgov1993,Dolgov2008,Blinnikov2016}, their model requires ad-hoc interactions and can give only a qualitative discussion.
On the other hand, our scenario is naturally described by the MSSM interactions and the observables are evaluated analytically.
Since the scenario requires no curvature perturbation, stringent constraint form $\mu$-distortion and PTA experiments are completely absent. 
Furthermore, the dark matter abundance in the current Universe is simultaneously explained by the non-topological solitons formed after AD baryogenesis, called Q-balls~\cite{Coleman1985,Enqvist1998,Enqvist1999,Kasuya2000g,Kasuya2000,Kasuya2001}. 
Consequently, the LIGO PBHs and dark matter are simultaneously explained in this model, that is, {\it cogenerated}
\footnote{Here we remark our scenario is different from the existing case where the dark matter abundance is explained by the LIGO PBHs themselves~\cite{Bird2016,Clesse2016,Kashlinsky:2016sdv}}.          

\section{Inhomogeneous AD baryogenesis}
First, let us consider the generation of the inhomogeneous baryon asymmetry, that is, the production of the HBBs. 
Although the mechanism is based on the AD baryogenesis as we mentioned before, we put two assumptions: 
(i) During inflation the AD-field has a positive Hubble induced mass, while it has negative one after inflation.
(ii) Just after inflation, the temperature of the decay products of the inflaton $T$ overcomes the Hubble parameter $H$. 
Although these assumptions seem to be somewhat unconventional, such a situation is conceivable in general\footnote{The coefficients of the Hubble induced mass term both during and after inflation are assumed to be negative in the context of the usual AD mechanism. In the supergravity-based inflation models, however, they are generally independent and take different values \cite{Kamada:2014qja,Kamada:2015iga,Hasegawa:2017rlz}.}.
Under these assumptions, the scalar potential for the AD-field~$\phi=\varphi e^{i\theta}$ is given by
\begin{align}\nonumber
 &V(\phi)=\\
&\begin{cases}
    (m_\phi^2+c_IH^2)|\phi|^2+V_{\rm NR},  &({\rm during~inflation}) \\
   (m_\phi^2-c_MH^2)|\phi|^2+V_{\rm NR}+V_{\rm T}(\phi),& ({\rm after~inflation})
  \end{cases}
\end{align}
where $c_I,~c_M$ are dimensionless positive constants, $m_\phi$ is the soft SUSY breaking mass for the AD-field ($\sim m_{3/2}$: gravitino mass) and $V_{\rm NR}$ denotes the non-renormalizable contributions given by 
\begin{align}
V_{\rm NR}=\left(\lambda a_M\frac{m_{3/2}\phi^n}{n\Mpl^{n-3}}+\hc\right)+\lambda^2\frac{|\phi|^{2(n-1)}}{\Mpl^{2(n-3)}},
\end{align}
where $\lambda, a_M$ are dimensionless constants. The integer $n~(\geq 4)$ is determined by specifying the MSSM flat direction. 
$V_{\rm T}$ is the thermal potential for the AD-field induced by the thermalized decay product of the inflaton and written as
\begin{align}\label{TP}
 V_{T}(\phi)&=\begin{cases}
    c_kf_k^2T^2|\phi|^2, & f_k|\phi|\ltsim T, \\
   a_g\alpha_s^2T^4\ln\left(\frac{|\phi|^2}{T^2}\right),& |\phi|\gtsim T,
  \end{cases}
\end{align}
where $c_k, f_k, a_g, \alpha_s$ are $\mathcal{O}(1)$ parameters relevant to the couplings of the AD-field to the thermal bath. 

\begin{figure}[t]
   \centering
   \includegraphics[width=85mm]{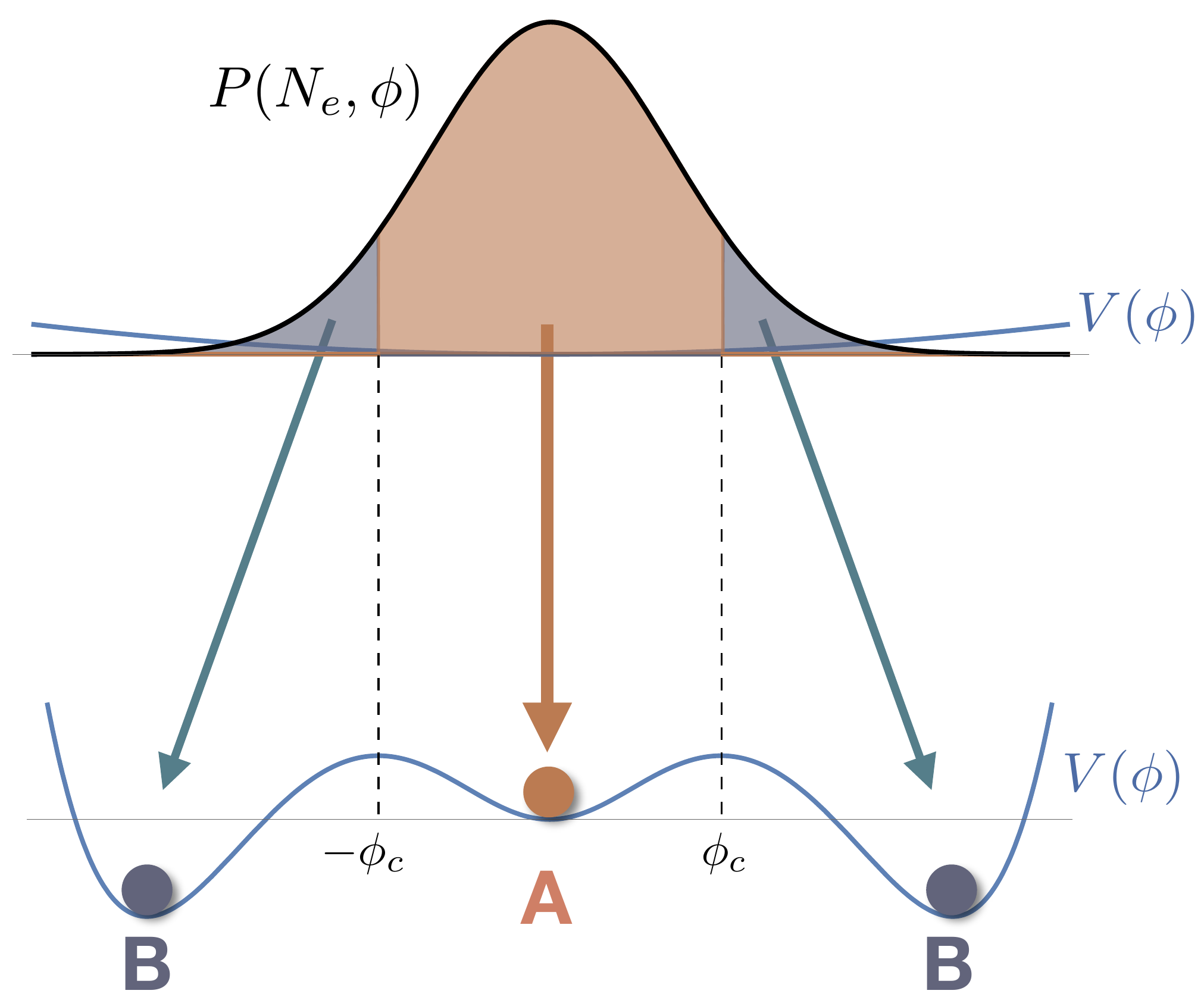}
   \caption{The schematic view of the bubble formation. The AD-field diffuses in the complex plane until the end of the inflation~(upper side). Just after inflation, the two vacuums A and B appear due to the thermal potential and the negative Hubble induced mass~(lower side). If $|\phi|>\phi_c$ in some patches, $\phi$ rolls down to the vacuum B. On the other hand, if $|\phi|<\phi_c$, $\phi$ rolls down to the vacuum A} 
   \label{fig:hbb}
\end{figure}

We can see this setting has a significant feature; \textit{multi-vacua appears after inflation}.
During inflation the potential has a minimum at the origin ($\phi=0$) by the assumption~(i).
After inflation, as usual,  the AD-field has a vacuum with a non-vanishing vacuum expectation value (VEV) $\varphi(t)\simeq(H(t)\Mpl^{n-3}/\lambda)^{1/(n-2)}$, due to the negative Hubble induced mass. 
We name this vacuum ``B".
In addition, around the origin $\varphi\lesssim f_k^{-1}T$, the thermal mass overcomes the negative Hubble induced mass because of the assumption~(ii), hence the second vacuum $\phi=0$ appears. 
We name this new vacuum ``A''. 
The shape of the potential is like a ``dented" Mexican hat as shown in the lower side of the Fig.\ref{fig:hbb}. 
The critical point between two vacuums lies at $\varphi_c(t)\simeq T(t)^2/H(t)$. 
Because the produced baryon number density is proportional to the value of $\varphi^2(t)$ at $H\simeq m_\phi$, substantial baryon asymmetry is produced at the vacuum B.
On the other hand, no baryon asymmetry is generated in the vacuum A. 

Let us describe the dynamics of the AD-field in the scenario. 
We show the schematic view of the dynamics in Fig.\ref{fig:hbb}.
During inflation, as mentioned above, the AD-field has positive mass and locates at the origin classically. 
However, the AD-field acquires quantum fluctuations during inflation. 
Therefore, IR modes of the AD-field ($=$ coarse-grained AD-field over local Hubble patches ) diffuse in the complex plane as the universe expands and $\phi$ takes different values in different Hubble patches~\cite{Vilenkin1982,Starobinsky1982,Linde1982}. 
After inflation, the shape of the potential is deformed to the ``dented" Mexican hat. 
Then, the AD-field rolls down to either of the two vacua A and B classically, and separate universe is realized.\footnote{At the moment, topological defects associated with spontaneous U(1) symmetry, could be formed. However, since the separated universe converges to the vacuum A after the AD mechanism, they immediately decay and could not cause cosmological problems.} 
If the AD-field takes a value $\varphi<\varphi_c$ in some patches at the end of inflation, $\varphi$ rolls down to the vacuum A and no baryon asymmetry is produced. 
On the other hand, if $\varphi>\varphi_c$, $\varphi$ rolls down to the vacuum B where the AD baryogenesis will occur. 
At the time $H\sim m$, vacuum B disappears due to the soft SUSY breaking mass and the AD field in the vacuum B start to oscillate around the vacuum A producing the baryon asymmetry.
As a result, the separate universe converses to the vacuum A and the difference in their path in the field space are reflected in the baryon asymmetry.
In the aim of the formation of the PBHs, we assume the baryon asymmetry produced via the phase B is very large such as $n_B^{\rm in}/s\sim1$\footnote{Such a large baryon asymmetry is naturally obtained by the AD-field with $n>6$.} and the regions of the phase B after inflation are very rare. 
Such highly-baryon asymmetric bubbles are called as high-baryon bubbles~(HBBs). 
Since in the rest of the universe, namely the phase A, the baryon asymmetry is not generated\footnote{One may consider the possibility that the HBBs account for observed baryon asymmetry. 
However, in high baryon regions, Big Bang Nucleosynthesis can not explain the observed abundances of the light elements. 
}, we have to prepare another mechanism which realizes the observed baryon asymmetry $n_B^{\rm ob}/s\sim10^{-10}$.

In the following discussions, we assume the instant thermalization of the decay products of the inflaton and evaluate their temperature $T(t)$ as
\begin{align}
T(t)\simeq(T_R^2H(t)\Mpl)^{1/4},
\end{align}
where $T_R$ is the reheating temperature. 
Then, the condition for the existence of the new vacuum, that is, assumption~(2) is translated as\footnote{If the condition is satisfied just after inflation, it holds at least until reheating completes.}
\begin{align}\label{Delta}
\Delta\equiv\frac{T_R^2\Mpl}{H_I^3}>1.
\end{align}
The critical point $\varphi_c(t)$ just after inflation is also rewritten as
\begin{align}
\varphi_c(t_e)\equiv\varphi_c=\Delta^{1/2}H_I,
\end{align}
where we set the $\mathcal{O}(1)$ parameters in Eq.~(\ref{TP}) as unity and treat the Hubble parameter during inflation $H_I$ as constant for simplicity.
\section{Distribution of HBBs}
Next, let we discuss the distribution of the HBBs. 
The AD-field with a positive mass $c_IH^2$ acquires the quantum fluctuations during inflation and its IR modes exhibit the Gaussian distribution~\cite{Vilenkin1982,Starobinsky1982,Linde1982} with a time-dependent variance
\begin{align}\label{qf}
\sigma^2(N)\equiv\langle\delta\phi^2(N)\rangle=\left(\frac{H_I}{2\pi}\right)^2(1-e^{-c'_IN})/c'_I, 
\end{align}   
where we define $N\equiv \ln(a/a_i),~c'_I\equiv(2/3)c_I$. 
The probability distribution function for the AD-field $\phi$ evaluated at $N$ is given by
\begin{align}
P(N,\phi)=\frac{e^{-\frac{\varphi^2}{2\sigma^2(N)}}}{2\pi\sigma^2(N)}.
\end{align}  
We can estimate the population of the HBBs by this stochastic discussion. 
Since the patches with $\varphi>\varphi_c$ are to be the HBBs after inflation, we call such patches also HBBs. The physical volume of the HBBs at certain $N$ is evaluated as
\begin{align}\label{frac}
V_{\rm B}(N)=V(N)\int_{\varphi>\varphi_c}P(N,\phi)d\phi\equiv V(N)f_B(N).
\end{align}   
Here we represent the physical volume of the Universe at $N$ as $V(N)\sim r_H^3e^{3N}$, where $r_H$ is the Hubble radius. 
$f_B$ denotes the volume fraction of the HBBs. 
The creation rate of the HBBs is obtained by differentiating $V_B(N)$ with respect to $N$:
\begin{align}
\frac{dV_{\rm B}(N)}{dN}=3V_{\rm B}(N)+V(N)\int_{\varphi>\varphi_c}\frac{dP(N,\phi)}{dN}d\phi.
\end{align}   
We can see that the first term represents the growth of the HBBs due to the cosmic expansion. The second term represents nothing but the creation of the HBBs at $N$. 
Therefore, the fraction of the HBBs formed at $N$ evaluated at the inflation end $N_e$ is
\begin{align}
\beta_B(N)=\frac{d}{dN}f_B(N)=\int_{\varphi>\varphi_c}\frac{dP(N,\phi)}{dN}d\phi.
\end{align}   
The result does not depend on $N_e$ because all the HBBs expand with same rate, and we can consistently reproduce Eq.(\ref{frac}) by integrating over all HBBs $(0<N<N_e)$. 
The integration over $\phi$ in Eq.(\ref{frac}) is straightforward and we can obtain the explicit form for $f_B$ (and so $\beta_B(N)$) as
\begin{align}
f_B(N)&=\int_0^{2\pi}d\theta\int^\infty_{\varphi_c}\varphi \frac{e^{-\frac{\varphi^2}{2\sigma^2(N)}}}{2\pi\sigma^2(N)}d\varphi=e^{-\frac{2\pi^2\Delta}{\tilde{\sigma}^2(N)}},
\end{align}   
where we define $\tilde{\sigma}^2(N)\equiv(1-e^{-c'_IN})/c'_I $. Therefore, surprisingly, the distribution of the HBBs created at $N$ is represented only by $c_I$ and the parameter $\Delta$ defined in Eq.(\ref{Delta}).

For later convenience, let us relate the size of the HBBs to the horizon mass $M_H$ evaluated at the time when the scale re-enters the horizon. Because the HBBs created at $N$ have the size of the Hubble horizon ($\sim H_I^{-1}$), we can relate the number of $e$-foldings $N$ with $M_H$ as
\begin{align}\label{N}
N(M_H)\simeq-\frac{1}{2}\ln\frac{M_H}{M_\odot}+21.5+N_{\rm CMB},
\end{align} 
where $M_\odot$ is the solar mass, $N_{\rm CMB}$ is the number of $e$-foldings when the pivot scale exits the horizon, and we used $g_*=10.75$ and the CMB pivot scale $k_*=0.002{\rm Mpc}^{-1}$. It is also convenient to relate the temperature $T$ at horizon crossing to the horizon mass as
\begin{align}\label{T}
T(M_H)=434{\rm MeV}\left(\frac{M_H}{M_\odot}\right)^{-1/2}.
\end{align} 

\section{PBH formation}
In this section, we describe how the HBBs form PBHs.
After inflation, the energy density inside and outside the HBB are almost same because the oscillation of the inflaton dominates the universe. However, the difference in the baryon asymmetry leads to the large density contrast in the late-time universe. 
   	
Inside the HBBs, the AD-field has a non-vanishing VEV and baryon asymmetry is generated at $H\sim m_\phi$ due to the AD mechanism. 
The coherent oscillation of the AD-field, however, is usually spatially unstable and fragments to the localized lumps, called Q-balls.
Q-ball is a configuration of the complex scalar which minimizes the energy under the fixed $U(1)$ charge ($=$ baryon number).

In this scenario, we consider the case the Q-balls are stable and almost all baryon charges are captured in the Q-balls. It is known that such a situation is naturally realized in the gauge-mediated SUSY~breaking scenario~\cite{Kasuya2001,Kasuya2000g}. We represent the abundance of the Q-balls formed after AD baryogenesis inside the HBBs as $\rho_Q^{\rm in}/s$\footnote{
The baryon asymmetry in the HBB depends on the initial value of the phase direction, which is generally different among HBBs. However, the Q-ball energy density is almost the same among HBBs because both Q-balls and anti-Q-balls are produced so that the total Q-ball density is equal to that of the AD field. Furthermore, we can also realize the same baryon density by introducing the Hubble induced A-term.
}.
Then, the density contrast of the HBB is
\begin{align}\label{delta}
\delta\equiv\frac{\rho^{\rm in}-\bar{\rho}}{\bar{\rho}}=\frac{\rho_Q^{\rm in}}{(\pi^2/30)g_*T^4}=\frac{4}{3T}\frac{\rho_Q^{\rm in}}{s},
\end{align}  
where $\rho^{\rm in},~\bar{\rho}$ denote the energy density inside and outside the HBBs, respectively. Since the Q-balls behave as pressure-less dust, their energy eventually dominates the HBBs. 
Defining the critical value of the density contrast for the gravitational collapse as $\delta_c$, we can see that PBHs start to be formed at
\begin{align}
T_c=\frac{4}{3}\delta_c^{-1}\frac{\rho_Q^{\rm in}}{s}.
\end{align} 
Here only the HBBs which are larger than the horizon scale at $T_c$ can gravitationally collapse into the PBHs. On the other hand, smaller HBBs can not form PBHs, but form the self-gravitational systems of the Q-balls which contribute to the dark matter abundance. 
The PBH formation from the over-density of the Q-balls has been studied in a different cosmological context \cite{Cotner:2016cvr,Cotner:2017tir}.

Here we comment on the equation of the state $p=w\rho$ inside the HBBs and the value of $\delta_c$. 
Before the Q-ball-radiation equality, we simply regard HBBs as almost radiation-dominated ($w\simeq1/3$) and use $\delta_c\simeq0.4$~\cite{Harada2013}.  On the other hand, after the Q-ball-radiation equality, we have to adopt the value of $\delta_c$ for $0<w<1/3$, which varies with time as well as $w$. However, soon after $T_c$, $\delta$ becomes larger than unity and hence without large errors we can assume that the HBBs collapse to the PBHs when they re-enter the horizon after $T_c$.

Then, the fraction of the PBHs with mass $M_{\rm PBH}$ is written as
\begin{align}\label{fr}
\beta_{\rm PBH}(M_{\rm PBH})=\beta_B(M_{\rm PBH})\theta(M_{\rm PBH}-M_c).
\end{align} 
Here we assume $M_{\rm PBH}\simeq M_H$ for simplicity. $M_c$ is a horizon mass at $T_c$. Hence, $M_c$ works as a ``cut-off" for the formation of the smaller PBHs.

\section{PBH abundance}

\begin{figure}[t]
   \centering
   \includegraphics[width=80mm]{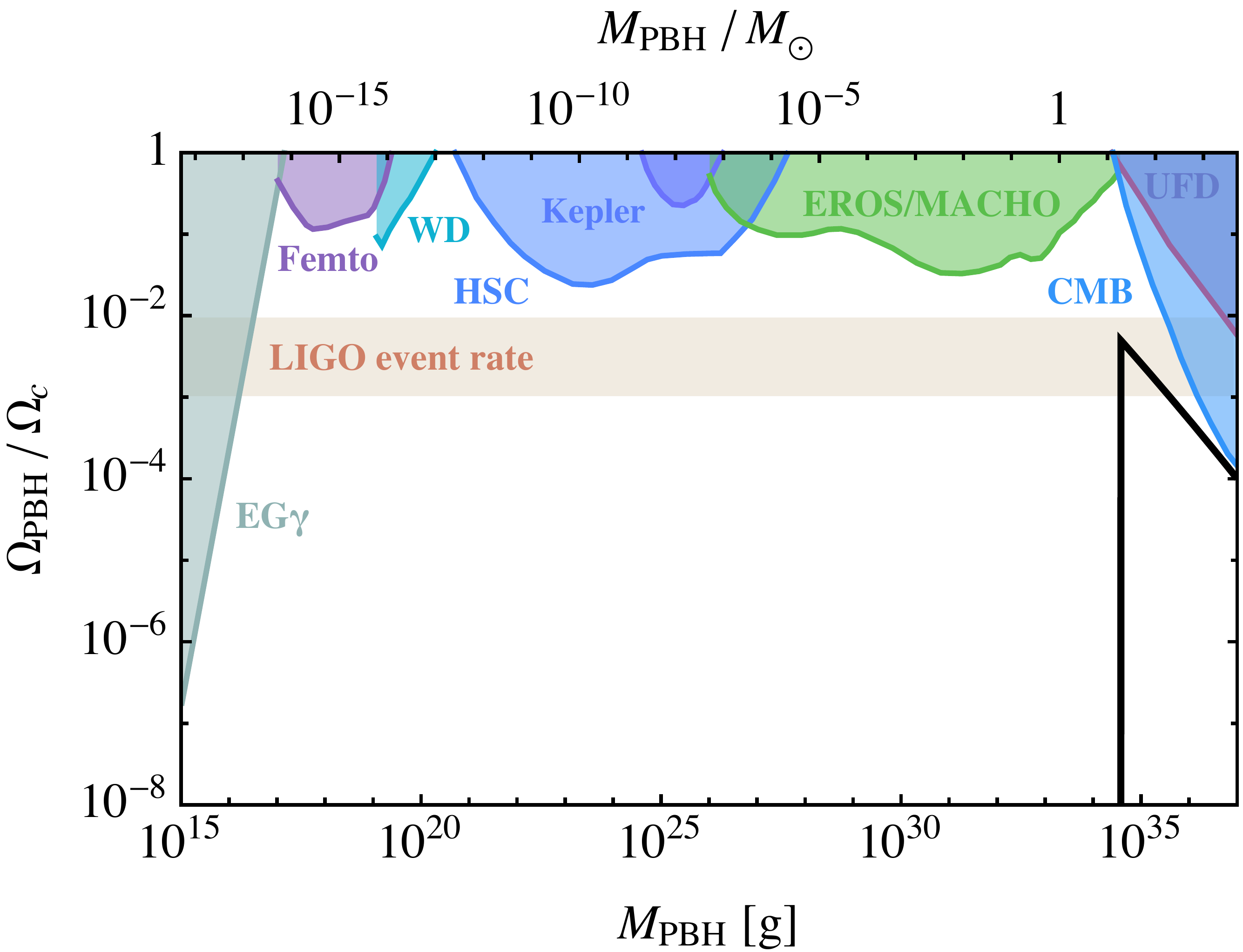}
   \caption{We show the PBH abundance for $(c_I,\Delta,N_{\rm CMB})=(0.046,19,10)$ and the observational constraints, The shaded regions are excluded by extragalactic gamma rays from Hawking radiation (EG$\gamma$)~\cite{Carr2009}, femtolensing of known gamma ray bursts (Femto)~\cite{Barnacka2012}, white dwarfs existing in our local galaxy (WD)~\cite{Graham2015}, microlensing search with Subaru Hyper Suprime-Cam (HSC)~\cite{Niikura2017}, Kepler micro/millilensing (Kepler)~\cite{Griest2013}, EROS/MACHO microlensing (EROS/MACHO)~\cite{Tisserand2007}, dynamical heating of ultra faint dwarf galaxies (UFD)~\cite{Brandt2016}, and accretion constraints from CMB (CMB)~\cite{Ali-Haimoud20172}. 
} 
   \label{fig:c}
\end{figure}

Let us calculate the present abundance of the PBHs. Since PBHs behave as matter, one can estimate the abundance of the PBHs with mass $M_{\rm PBH}$ over logarithmic mass interval $d\ln M_{\rm PBH}$ as  
\begin{align}\nonumber
&\frac{\Omega_{\rm PBH}(M_{\rm PBH})}{\Omega_c}\\\nonumber
&\simeq\left.\frac{\rho_{\rm PBH}}{\rho_m}\right|_{\rm eq}\frac{\Omega_m}{\Omega_c}=\frac{\Omega_m}{\Omega_c}\frac{T(M_{\rm PBH})}{T_{\rm eq}}\beta_{\rm PBH}(M_{\rm PBH})\\\label{pa}
&\simeq\left(\frac{\beta_{\rm PBH}(M_{\rm PBH})}{1.6\times10^{-9}}\right)\left(\frac{\Omega_ch^2}{0.12}\right)^{-1}\left(\frac{M_{\rm PBH}}{M_\odot}\right)^{-1/2},
\end{align}
where $\Omega_c$ and $\Omega_m$ are the present density function of the dark matter and matter, respectively. Here we use the latest Planck result $\Omega_ch^2\simeq0.12$~\cite{PlanckCollaboration2015a}. $T(M_{\rm PBH})$ and $T_{\rm eq}$ are the temperatures at the formation of the PBHs with mass $M_{\rm PBH}$ and the matter-radiation equality, respectively. We show the PBH abundance Eq.(\ref{pa}) and observational constraints in Fig.~\ref{fig:c}. 
Here, restricting our interest to the PBHs inferred from LIGO events, we set $T_c=200{\rm MeV}$ which corresponds to $M_{c}\simeq \mathcal{O}(10)M_\odot$. 
We can see that due to the cut-off $M_c$, there exists a peak-like ``edge" whose mass $\sim\mathcal{O}(10)M_\odot$ and the abundance $\Omega_{\rm PBH}/\Omega_c\sim\mathcal{O}(10^{-2})-\mathcal{O}(10^{-3})$ can explain the merger rate~\cite{Sasaki2016} for the LIGO events. Also we have numerically estimated that 
\begin{align}\label{er}
f_B\gtsim\mathcal{O}(10^{-10})
\end{align}
is required to make the peak to reach at least $\Omega_{\rm PBH}/\Omega_c\sim\mathcal{O}(10^{-3})$. 

On the other hand, the Q-balls in the smaller HBBs, which did not collapse to PBHs, make a contribution to the dark matter abundance. 
Taking into account that only a small fraction of HBBs collapse to the PBHs, the abundance of these residual Q-balls is estimated as
\begin{align}\nonumber
\frac{\rho_Q}{s}&\simeq f_B\frac{\rho_Q^{\rm in}}{s}\\
&=4.4\times10^{-10}{\rm GeV}\left(\frac{T_c}{200{\rm MeV}}\right)\left(\frac{\delta_c}{0.4}\right)\left(\frac{f_B}{7.3\times10^{-9}}\right).
\end{align}
Comparing with Eq.(\ref{er}), we can conclude the residual Q-balls must contribute to the dark matter abundance more than $\mathcal{O}(10)\%$ to explain the LIGO event rate. 
Surprisingly, the LIGO PBHs and dark matter are simultaneously generated, namely, {\it cogenerated} in our scenario. 
Actually, the parameter choice $(c_I,\Delta,N_{\rm CMB})=(0.046,19,10)$ we made in the Fig.\ref{fig:c} explains the all dark matter by the residual Q-balls. 

Here we make a remark about the lightest SUSY particle (LSP). 
In the gauge-mediated SUSY breaking which predicts stable Q-balls, the LSP is gravitino with mass less than $1$~GeV. 
The LSP gravitino is stable and its abundance is roughly proportional to the reheating temperature. 
In the present scenario we assume that the reheating temperature is low enough ($T_R \lesssim 10^6$~GeV) for the gravitino to give a negligible contribution to the dark matter~\cite{Moroi:1993mb}.
In this case, the condition Eq.~(\ref{Delta}) is satisfied if the Hubble parameter during inflation is small ($H_I \lesssim 10^{10}$~GeV).

\section{Conclusions and Discussions}

In this letter, we have proposed a novel scenario which simultaneously explains the LIGO GW events and dark matter in the MSSM. 
The scenario is based on the generalized version of the AD mechanism which produces the highly localized baryon asymmetry called HBBs.
If the Q-balls created after AD mechanism are stable, HBBs eventually become overdense due to the additional energy contribution from the Q-balls and gravitationally collapse to the PBHs at a certain time. 
We have showed that the scenario predicts $\mathcal{O}(10)M_\odot$ PBHs with abundance $\Omega_{\rm PBH}/\Omega_c\sim\mathcal{O}(10^{-2})-\mathcal{O}(10^{-3})$, which account for the LIGO events. 
We stress that the scenario generically evades the stringent constraints from PTA experiments and $\mu$-distortion because the averaged amplitude of small scale curvature perturbations is sufficiently small. 
Furthermore, Q-balls inside the residual HBBs, which are too small to gravitationally collapse, can account for all dark matter. 
These Q-balls are considered to form self-gravitational systems in every HBB and would have interesting implications in late-time cosmology.

We comment on the case without stable Q-balls, that is, Q-balls decay before their domination or are not formed after AD baryogenesis. 
Actually, PBHs are also created in this case. 
This is because the large baryon number density in the HBBs is transformed to matter energy density due to the QCD phase transition. 
While the evaluation of the PBH abundance is same as Eq.(\ref{fr}), the cut-off for the lighter PBHs $M_c$, at which the peak-like ``edge" locates, must be determined as
\begin{align}
M_c\simeq M_{\rm QCD}\simeq 19M_\odot.
\end{align}
Interestingly, the mass scale which corresponds to the LIGO events is {\it predicted} by the QCD dynamics. We will study this case in more detail in future~\cite{Hasegawa20172}. 

Finally, we mention the existence of the SMBHs. It is seen from Fig.\ref{fig:c} that a significant amount of SMBHs are created in a wide mass range. 
Although they successfully evade the constraints from the accretion and $\mu$-distortion\footnote{There have been some attempts to explain the SMBHs by the PBHs evading the $\mu$-distortion constraints \cite{Kohri:2014lza,Nakama:2016kfq,Nakama:2017xvq}. }, their abundance is much greater than one in every comoving volume of $1{\rm Gpc}^3$. 
Such excessive SMBHs are originated from HBBs formed near the beginning of the inflation\footnote{On the other hand, the ref.\cite{Nakama:2017xvq} suggests the more abundant SMBHs such as $\Omega_{\rm SMBH}/\Omega_c\sim10^{-4}$, which is compatible with our result.}. 
The SMBHs are suppressed if the IR modes start to diffuse just before the scale corresponding to $M_c$ exits the horizon. 
Then the HBBs much heavier than $M_c$ are not created in principal and the peak at $M_c$ becomes sharper. 
This can be realized in double inflation; during the first inflation the AD-field is stabilized at the origin due to the large Hubble induced mass ($c_I\gg1$), while the AD field obtains smaller mass ($c_I \ll 1$) and starts to diffuse during the second inflation. We will also discuss this case in ref.\cite{Hasegawa20172}
\section*{Acknowledgement}
We would like to thank Kenta Ando, Jeong-Pyong Hong, Masahiro Ibe, Keisuke Inomata and Eisuke Sonomoto for helpful comments. This work is supported by JSPS KAKENHI Grant Number 17H01131 (M. K.) and 17K05434 (M. K.), MEXT KAKENHI Grant Number 15H05889 (M. K.), JSPS Research Fellowship for Young Scientists Grant Number 17J07391 (F. H.) and also by the World Premier International Research Center Initiative (WPI), MEXT, Japan.


\bibliographystyle{apsrev4-1}
\bibliography{PBH}

\end{document}